# Observation of Weyl nodes in robust type-II Weyl semimetal WP$_2$

M.-Y. Yao[1,*], N. Xu[1,2,3,†], Q. Wu[2,4], G. Autès[2,4], N. Kumar[5], V. N. Strocov[1], N. C. Plumb[1], M. Radovic[1], O. V. Yazyev[2,4], C. Felser[5], J. Mesot[1,2,6], M. Shi[1,‡]

[1]*Swiss Light Source, Paul Scherrer Institut, CH-5232 Villigen PSI, Switzerland*

[2]*Institute of Physics, École Polytechnique Fédérale de Lausanne (EPFL), CH-1015 Lausanne, Switzerland*

[3]*Institute of Advanced Studies, Wuhan University, Wuhan 430072, China*

[4]*National Center for Computational Design and Discovery of Novel Materials MARVEL, École Polytechnique Fédérale de Lausanne (EPFL), CH-1015 Lausanne, Switzerland*

[5]*Max Planck Institute for Chemical Physics of Solids, 01187 Dresden, Germany*

[6]*Laboratory for Solid State Physics, ETH Zürich, CH-8093 Zürich, Switzerland*

\* E-mail: mengyu.yao@psi.ch
† E-mail: nxu@whu.edu.cn
‡ E-mail: ming.shi@psi.ch

**Distinct to type-I Weyl semimetals (WSMs) that host quasiparticles described by the Weyl equation, the energy dispersion of quasiparticles in type-II WSMs violates Lorentz invariance and the Weyl cones in the momentum space are tilted. Since it was proposed that type-II Weyl fermions could emerge from (W,Mo)Te$_2$ and (W,Mo)P$_2$ families of materials, a large numbers of experiments have been dedicated to unveil the possible manifestation of type-II WSM, e.g. the surface-state Fermi arcs. However, the interpretations of the experimental results are very controversial. Here, using angle-resolved photoemission spectroscopy supported by the first-principles calculations, we probe the tilted Weyl cone bands in the bulk electronic structure of WP$_2$ directly, which are at the origin of Fermi arcs at the surfaces and transport properties related to the chiral anomaly in type-II WSMs. Our results ascertain that due to the spin-orbit coupling the Weyl nodes originate from the splitting of 4-fold degenerate band-crossing points with Chern numbers $C = \pm 2$ induced by the crystal symmetries of WP$_2$, which is unique among all the discovered WSMs. Our finding also provides a guiding line to observe the chiral anomaly which could manifest in novel transport properties.**

In the past decade, realization of topological phases in solids has become an important subject in experimental condensed matter physics. Among numerous kinds of topologically nontrivial materials, the Weyl semimetals (WSMs) have attracted a great deal of attention [1-8]. In these materials the low-energy excitations (quasiparticles) near the band crossing points (Weyl nodes) behave as massless fermions with non-zero chirality. In theoretical considerations there are two types of WSMs in which the spin degeneracy of energy bands is lifted by breaking time-reversal or inversion symmetry. In type-I WSMs the low-energy excitations near the band crossing points can be described by the Weyl equation and the quasiparticles are isomorphic to the Weyl fermions predicted in particle physics [9]. In type-II WSMs, the Weyl nodes appear at the touching points of electron and hole pockets. The emerging quasiparticles violate Lorentz invariance and exhibit tilted cone-like low-energy band dispersion [10-12].

The concept of type-II WSM was brought forward by studying the topological properties of $WTe_2$ and $MoTe_2$ [13-24]. Great efforts were made to experimentally identify the spectroscopic evidence of the Fermi arcs, a consequence of the Weyl semimetal phase, as well as to explore the transport properties that could be associated with type-II Weyl fermion excitations. However, as the predicted Weyl nodes lie above the Fermi level ($E_F$), the interpretations of the experimental results are very controversial, and it remains unclear whether the observed surface states or the transport properties are a direct consequence of the presence of type-II Weyl cones in the bulk.

Recently, it was suggested that $WP_2$ and $MoP_2$ host robust type-II Weyl fermions with Weyl nodes appearing below $E_F$ [25]. Distinct to all other (type-I and type-II) WSMs, the Weyl nodes in (W,Mo)$P_2$ emerge from the splitting of 4-fold degenerate band-crossing points with Chern numbers $C = \pm 2$ induced by the crystal symmetries. These materials are semimetals that crystalize in the orthorhombic structure (Fig.1a), which belongs to non-symmorphic $Cmc2_1$ space group. The Brillouin zone (BZ) and the (0 1 -1) cleavage plane (as indicated by yellow plane) is shown in Fig. 1b. According to the theoretical prediction, if there were no spin-orbit coupling (SOC), hole pockets and electron pockets would have formed four touching points located in the $\Gamma$-$X$-$Y$ plane. Each touching point is a 4-fold degenerate linear band crossing point with Chern numbers $C = \pm 2$, and all the touching points are well separated from each other in momentum space. When SOC is taken into account, each 4-fold degenerate point splits into two 2-fold Weyl nodes with same chirality ($C = +1$ or $-1$) appearing at different binding energies. In total there are eight Weyl nodes in each BZ. A schematic band structure of two adjacent type-II Weyl nodes contains two touching points of a hole and an electron pocket, as shown in Fig. 1c. The adjacent Weyl nodes, W1 and W2 (or equivalently W1' and W2') have the same chirality ($C = +1$ or $-1$), while the Weyl nodes with opposite chirality (i.e. W1 and W1') are well separated in momentum space. Because of the large separation of the Weyl nodes with different chirality, two adjacent Weyl nodes cannot annihilate with each other under small perturbations, which ensures that the topologically non-trivial phase of the system is robust. According to the DFT calculation [25], the Weyl nodes in $WP_2$ remain unchanged under positive and negative strain ranging from +4% to -4%. In comparison, for $WTe_2$, the Weyl nodes with opposite chirality annihilate for negative strain beyond -0.5%. As a consequence, the Weyl semimetal phase in $MoP_2$ and $WP_2$ is expected to be robust against defects or structural distortions. To ascertain the theoretical prediction that $WP_2$ is a type-II WSM, it is crucial to unveil the Weyl nodes in the bulk electronic structure.

Combined with first-principles calculations, we present a comprehensive soft X-ray angle-resolved photoemission spectroscopy (SX-ARPES) investigation on the high-quality single-crystal $WP_2$. Compared to ARPES using ultraviolet light (VUV-ARPES), because photoelectrons have higher kinetic energies in SX-ARPES measurements, the photoelectron escape depth becomes much larger. The enlarged escape depth leads to increased bulk sensitivity because photoelectrons can emit from deeper bulk region below the surface. Furthermore, due to the Heisenberg uncertainty principle, increasing the photoelectron escape depth results in high intrinsic $k_z$ resolution as the $k_z$ broadening is inversely proportional to the escape depth [26]. Since the Weyl fermions are bulk originated, SX-ARPES is a more ideal tool to study the band structure. Although using VUV-APRES it has been shown that the bulk-band structure of $WP_2$ is electron-hole compensated, it is important to explore the touching points of electron and hole pockets of the bulk states [27]. Therefore, by taking advantage of bulk-sensitive SX-ARPES, our study not only provides mainly bulk and higher $k_z$ resolution of $WP_2$ electronic structure, but also for the first time, characterize the the Weyl nodes of $WP_2$. In agreement with density functional theory (DFT) calculations, our results reveal two pairs of Weyl nodes lying at different binding energies in the first BZ. The observation of the Weyl nodes, as well as the tilted cone-like dispersions in the vicinity of the nodal points, provides a compelling evidence that WP$_2$ is a robust type-II WSM with broken Lorentz invariance, and offer a playground for further investigation of this new type of Weyl fermions.

Single crystals of WP$_2$ were prepared by chemical vapor transport method. The ARPES experiments were carried out at the ADRESS beamline at Swiss Light Source, with a SPECS analyzer, and data were collected using circularly polarized light with an overall energy resolution of the order of 50 - 80 meV [28]. The photon energy is in the soft X-ray region (300 - 800 eV), which is highly sensitive to bulk states. The samples with a typical size of ~2×0.3×0.3 mm$^3$ were cleaved at 15 K in high vacuum chamber, with base vacuum better than 5x10$^{-11}$ Torr. The bulk electronic band structure was calculated within the density functional theory (DFT) framework using the generalized gradient approximation (GGA) as implemented in the QUANTUM-ESPRESSO software package. The electronic structure calculations were cross-validated with Vienna ab-initio simulation package (VASP) [29,30]. Spin-orbit coupling (SOC) is taken into account with the help of fully relativistic ultrasoft pseudopotentials. The calculations were carried out using an 8×8×5 *k*-point mesh and a planewave kinetic energy cutoff of 50 Ry for the wavefunctions.

We first perform core-level photoemission measurements, as shown in Fig. 1d. The ARPES spectrum of WP$_2$ acquired with *hv* = 650 eV shows the characteristic peaks of W and P elements, confirming the chemical composition of WP$_2$ samples. Next, we investigate the electronic structure of WP$_2$ by means of systematic ARPES measurements. Figure 2 shows the Fermi surface (FS) of WP$_2$, calculated using DFT and determined experimentally from ARPES measurements, as well as the band dispersions along the high-symmetry lines. The calculated FS is rather complex, there is an open spaghetti-like hole FS pocket located near the *X* point (Fig. 2a) and a closed bow-tie-like electron FS pocket located around the *Y* point. To experimentally verify the predictions of the electronic structure calculations, we have carried out ARPES on the (0 1 -1) cleavage surface of WP$_2$ (Fig. 2b). Figure 2c shows the ARPES intensity map at $E_F$ in the $k_x$-$k_y$ plane that contains the *Z* and *Y* points in the first BZ, as well as the Γ and *X* points in the second BZ, acquired with photon energy *hv* = 415 eV. In this $k_x$-$k_y$ plane the cross section of the bow-tie-like FS pocket appears a butterfly-shape electron pocket around the *Y* point, and the spaghetti-like

FS forms hole pockets around the $X$ points. Those features can be found on the calculated FS (Fig. 2d). It should be noted that the calculated FS in Fig. 2(d) takes into account the change in $k_z$ as a function of emission angle of the photoelectrons [31], in order to accurately reproducing the curved surface in momentum space, from which the ARPES data were acquired for a given photon energy. Probably due to the photon energy dependence of the matrix element for photoemission, the butterfly shaped electron pocket is only weakly observable at $h\nu = 415$ eV, while it is more clearly visible for $h\nu$ ranging from 370 eV to 390 eV (Fig. 2g). To show the agreement between ARPES data and calculated bands, we plot ARPES spectra along the $X$-$\Gamma$ and $Z$-$Y$ lines with calculated band dispersions overlaid on top (red lines) in Figs. 2e,f. Two bands cross $E_F$ forming a hole pocket around the $X$ point (Fig. 2e), corresponding to the hole pocket shown on the FS map in Fig. 2c. At the $Y$ point, an electron band touches $E_F$, leaving a very shallow electron pocket (Fig. 2f). The overall feature of the measured band structure is in good agreement with the calculation. By tuning $h\nu$ from 340 eV to 390 eV we obtained the FS map at different $k_z$ values (Fig. 2g). The evolution of band dispersion along $k_z$ verifies the 3D nature of the electronic structure, therefore confirms its bulk origination.

The most crucial step to justify that $WP_2$ is a type-II WSM is to identify the Weyl points predicted by first-principles calculations. The Weyl nodes are predicted to be located in the $\Gamma$-$X$-$Y$ plane (Fig. 3a), which can be accessed by using $h\nu = 360$ eV photons. Figure 3b shows five constant energy maps acquired with $h\nu = 360$ eV. From Fig. 3b, the electron and hole pockets are clearly identified, and their profiles are indicated by red and white dashed lines, respectively. At $E_F$, the electron and hole pockets are detached as shown in panel (i). With decreasing binding energy, the electron and hole pockets approach each other and were found to have a first touching point that characterize a type-II Weyl node at $E_B \sim -0.28$ eV, marked as W1 in the panel (ii). With further decreasing binding energy, the electron pocket continues to shrink, while the hole pocket expands. They are separated slightly in the panel (iii). At $E_B \sim -0.41$ eV, the pockets have another touching point, marked as W2 in the panel (iv). Finally, the electron pocket disappears, as shown in panel (v). To further confirm our assumption that the touching points in the panel (ii) and (iv) are the Weyl nodes, ARPES spectra along $k_x$ direction and through the touching points identified in the constant energy maps were acquired, in which we observed two band-crossing points formed by an electron and a hole band (Fig. 3c,e). In Fig. 3d,f, two ARPES spectra along $k_y$ direction were taken at the $k_x$ value of the crossing points shown in Fig. 3c,e, respectively, from which the touching points of the electron and hole pockets can be observed. The band structures near the crossing points are in good agreement with predicted Weyl nodes. The W1 and W2 nodes are located at ~ 0.28 and 0.41 eV below the $E_F$, respectively, which agrees well with the calculated values (0.340 and 0.471 eV) assuming a rigid shift of calculated bands upwards by ~ 60 meV. This energy difference could result from doping due to defects or other lattice imperfections [32]. Figure 3d,f show the ARPES spectra overlaid with the calculated band structure along the $k_y$ direction, the Weyl cones containing W1 and W2 are tilted in this direction. From the obtained electronic structure it can be seen clearly that the W1 and W2 Weyl node are not the local extrema along the $k_y$ direction, which is the key feature of type-II Weyl fermions and distinct from the Weyl nodes in type-I WSMs, e.g. TaAs and TaP. Therefore, we have identified the touching points between electron and hole pockets (Fig. 3b(ii, iv)) at the positions in energy and momentum space at which the DFT calculation predict the Weyl nodes.

Furthermore, WP$_2$ (and MoP$_2$) is unique compared to other WSMs because of the robust Weyl nodes. As shown in Fig. 3g, due to the sizable SOC in WP$_2$, two 2-fold Weyl nodes with same chirality ($C = +1$ or -1) split from a 4-fold degenerate point ($C = \pm 2$). As a result, Weyl nodes with different chirality are largely separated in momentum space. The Weyl nodes with opposite chirality (W1 and W1', W2 and W2') are well separated along the $k_x$ direction, the separation between W1 and W1' (W2 and W2') is as large as ~ 1/4 of the BZ dimensions in this direction. Therefore, two adjacent Weyl nodes cannot annihilate with each other under small perturbations, which ensures that the topologically non-trivial phase of the system is robust. This is different from the situation when the 4-fold degeneracy of a Dirac point that has $C = 0$ is lifted by breaking time-reversal or inversion symmetry, which leads to two 2-fold Weyl nodes with opposite chirality ($C = +1$ and -1).

With the presence of type-II Weyl fermions, negative magnetoresistance (MR) resulting chiral anomaly should be expected in WP$_2$. On the other hand, in recent transport measurement, the negative MR is not observed [33]. As predicted by theory, different from type-I WSM, the observation of negative MR in type-II WSM requires to apply electric field and magnetic field within the cone formed by electron and hole pocket [10]. Our study ascertains that the Weyl cones are tilted along $k_y$ direction for both W1 and W2, which would suggest that the chiral anomaly can be observed only when the *E*-field and *B*-field are both along the $k_y$ direction, which is perpendicular to the natural growth direction.

This work was supported by the NCCR-MARVEL funded by the Swiss National Science Foundation, and the Sino-Swiss Science and Technology Cooperation (Grant No. IZLCZ2-170075), the ERC Advanced Grant No. (742068) "TOPMAT", the National Natural Science Foundation of China (No. 11874047), the Ministry of Science and Technology of China (2018YFA0307000), the National Key R&D Program of China (Grant No. 2018FYA0305800), and "the Fundamental Research Funds for the Central Universities" (Grant No. 2042018kf-0030).

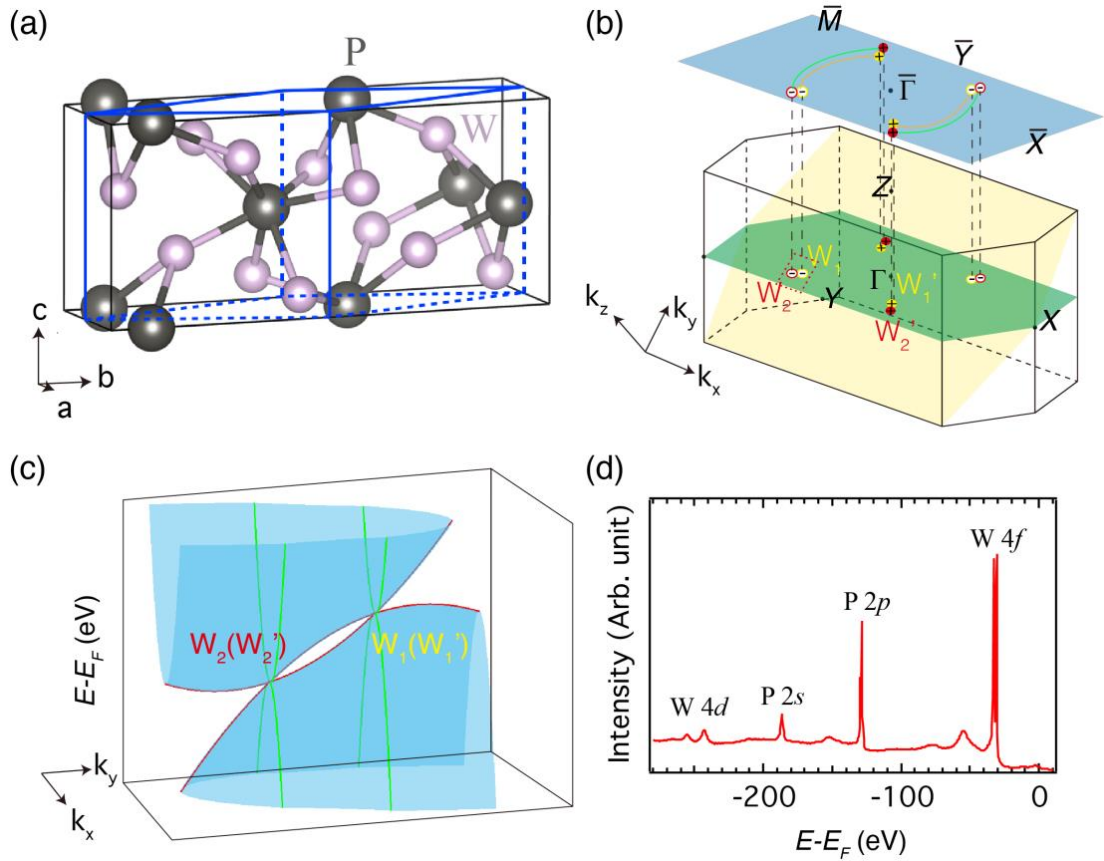

FIG. 1. (a) Crystal structure of WP2. (b) Bulk and surface BZ of WP2 with high-symmetry points labeled. The green plane represents the Γ-*X*-*Y* plane. The yellow plane represents the (0 1 -1) cleavage plane. In the momentum coordinate, $k_x$-$k_y$ is setup within the (0 1 -1)

plane. W1(W1') and W2 (W2') Weyl nodes are indicated by yellow and red open (closed) circles, respectively. The chirality of the Weyl nodes is indicated with + and − signs. The blue plane represents the 2D surface BZ. The orange and green lines schematically illustrate the possible Fermi arc dispersion connecting the Weyl nodes with opposite chirality. (c) Schematic band dispersions of type-II Weyl nodes in the $k_x$-$k_y$ plane of the momentum space. The region in momentum space is schematically indicated by red dashed lines in (b). (d) WP$_2$ core-level spectrum, obtained with $hv$ = 650 eV.

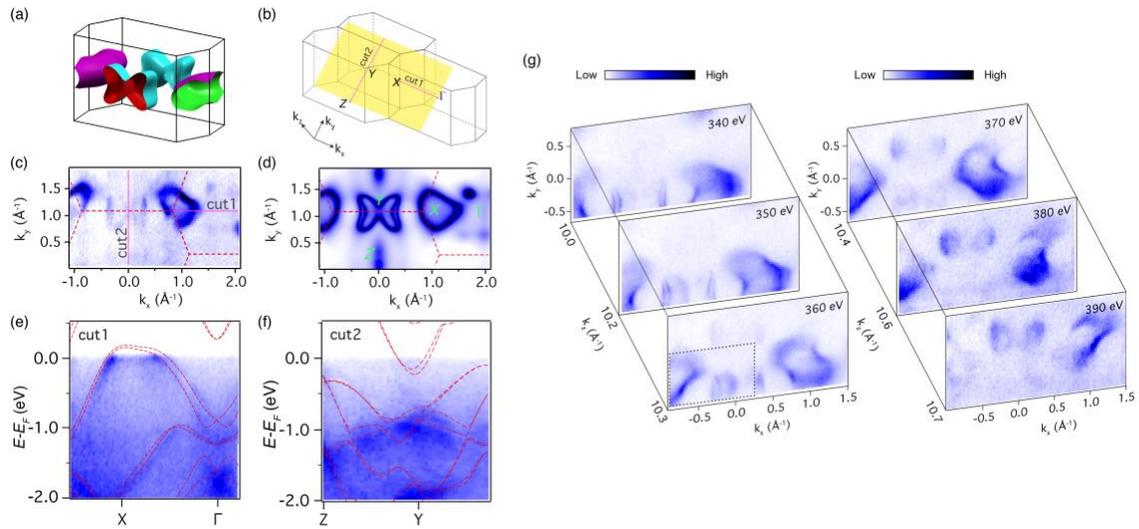

FIG. 2. (a) FSs of bulk bands from first-principles calculations. (b) BZ of WP$_2$ with the high-symmetry points indicated. The yellow plane represents the (0 1 -1) plane. (c-d) Constant energy maps at $E_F$ measured experimentally with $hv$ = 415 eV photons and obtained from first-principles calculations, respectively. (e) The ARPES intensity plots with calculated bands overlaid on top (red lines). The data were acquired with $hv$ = 415 eV along cut1. The calculated bands are rigid shifted to match the ARPES data. (f) Same as (e) but along cut2. (g) The FS map taken along $k_z$, obtained with $hv$ ranging from 340 eV to 390 eV.

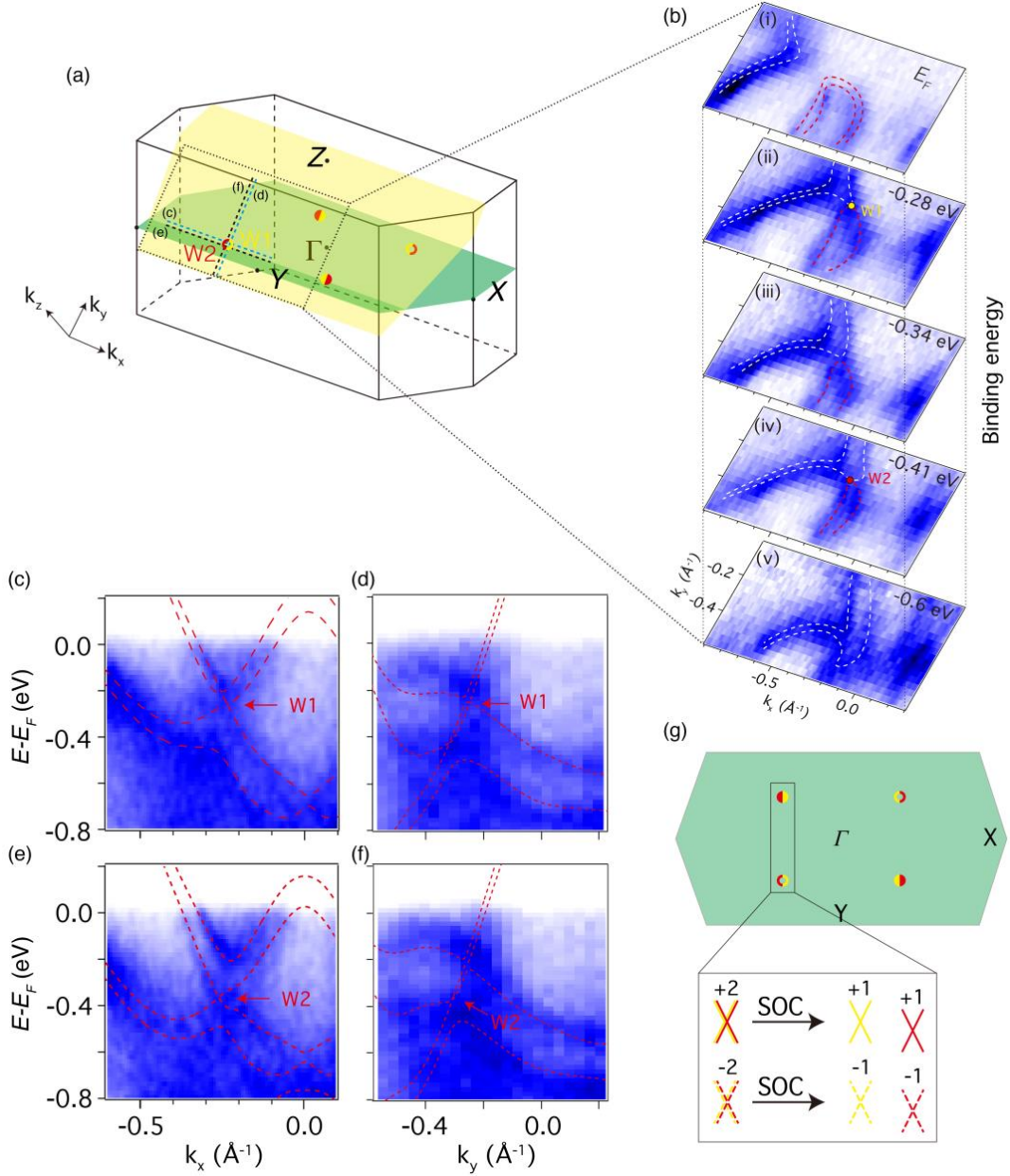

FIG. 3. (a) Theoretical predicted Weyl nodes in 3D BZ. The yellow plane represents the plane in k-space that across the Weyl nodes. (b) Series constant energy maps obtained with $h\nu = 360$ eV. The responding region is indicated in the 360-eV spectrum of Fig. 2(g). Panel (i - v), $E_B = E_F$, -0.28, -0.34, -0.41 and -0.6 eV, respectively. The red and white dashed lines are guide to eye, showing the profiles of the electron and hole pockets, respectively. (c-d) The ARPES spectra, obtained with $h\nu = 360$ eV, and the corresponding calculated band structures along the $k_x$ and $k_y$ directions across the W1 Weyl nodes, as marked by the dashed lines in (a). (e-f) The same ARPES spectra and calculated bands for the W2 Weyl point. (g)

(Upper panel) Four 4-fold degenerate points ($C = \pm 2$) located in the $\Gamma$-$X$-$Y$ plane in the WP$_2$ bulk BZ. (Lower panel) When SOC is taken into account, the 4-fold degenerate points ($C = \pm 2$) split into two 2-fold Weyl nodes with the same chirality ($C = +1$ or $-1$). Yellow and red color in the upper panel represent the Weyl nodes located at different binding energies.